# A New Proposed Cost Model for List Accessing Problem using Buffering

### Rakesh Mohanty
Dept. of Comp. Sc. & Engg.
Indian Institute of Technology
Madras, Chennai, India

### Seetaya Bhoi
Dept. of Comp. Sc. & Engg
Veer Surendra Sai Institute of
Technology
Burla, Orissa, India

### Sasmita Tripathy
Dept. of Comp. Sc. & Engg
Veer Surendra Sai Institute of
Technology
Burla, Orissa, India

## ABSTRACT
There are many existing well known cost models for the list accessing problem. The standard cost model developed by Sleator and Tarjan is most widely used. In this paper, we have made a comprehensive study of the existing cost models and proposed a new cost model for the list accessing problem. In our proposed cost model, for calculating the processing cost of request sequence using a singly linked list, we consider the access cost, matching cost and replacement cost. The cost of processing a request sequence is the sum of access cost, matching cost and replacement cost. We have proposed a novel method for processing the request sequence which does not consider the rearrangement of the list and uses the concept of buffering, matching, look ahead and flag bit.

## 1.INTRODUCTION
The list accessing problem involves maintaining and organizing a dictionary as a linear list. A dictionary is an abstract data type that stores and maintains a set of elements and supports the operations access, insert, and delete. For accessing an element, the list is traversed from the start of the list until the requested element is found. The insertion operation involves addition of an element at the end of the list. An element is deleted by first searching for the element and then removing it. As insertion and deletion of an element can be considered as a special case of access operation, therefore we can consider the access operation only for maintaining and organizing the dictionary.

### 1.1  Problem Statement
In list accessing problem, an unsorted linear list L of l distinct elements is given as input along with a finite sequence of requests of size n such that ($n \geq l$). Here each input request is an access operation. The list accessing algorithm takes an unsorted linear list and a request sequence as input and serves the requests in order of their arrival. A request is said to be served, when an access operation is performed on the requested element by incurring some access cost. Accessing an element 'x' at position 'i' from the front of the list costS 'i'. Our goal is to reduce the total access cost while serving a request sequence on the list.

### 1.2  Applications
The list accessing technique is extensively used in storing and maintaining small dictionaries. One important application of list accessing technique is data compression. Other applications include computing point maxima and convex hulls in computational geometry, organizing the list of identifiers maintained by a compiler and resolving collisions in a hash table. The list accessing problem is also of significant interest in the contest of self organizing data structures.

### 1.3  Related Work
The list accessing techniques were initiated by the pioneering work of McCabe in 1965[1]. He investigated the problem of maintaining a sequential file and developed two algorithms Move To Front and Transpose. Sleator and Tarjan in 1985 proposed a standard full cost model for the list accessing problem[2], which is the most widely used cost model. It involves free exchanges and paid exchanges to rearrange the input list. The partial cost model[3] assigns the cost by counting the number of comparisons. A comprehensive survey of List Accessing Problem along with various cost models has been done in [4], [5], [6], [ 7], [8].

### 1.4  Our Contribution
In this paper, we have made a study of different existing cost models for list accessing problem and proposed a new cost model. The uniqueness of our proposed cost model is that it assigns the cost using the concept of buffering, the look ahead and matching. We have proposed an algorithm which involves above concepts and calculate the cost by using our proposed model.  This algorithm does not involve the rearrangement of the input list. We have also analysed the performance of proposed cost model by using the developed algorithm.

### 1.5  Organization of Paper
   This paper is organised as follows. Section II contains description of well known cost models and some well known list accessing algorithms. Section III presents our new proposed cost model and evaluation of access cost using our model for the list accessing problem.. Section IV   provides the concluding remarks and scope of future research work.

## 2.  PRELIMINARIES

### 2.1  Cost Models
When an element is accessed in the linear list, a cost is assigned to it. This assignment of cost is defined by different cost models. There are various cost models for the list accessing problem using singly linked list data structure such as full cot model, partial cost model, paid exchange model etc. A start pointer is pointed to the beginning of the list and the list is to be traversed from the start pointer till the requested element is found in the list. The two most widely used cost models for list accessing problem using singly linked list are Full cost Model and Partial cost Model. These models assume that after an item has been requested, it may be moved free of charge closer to the front of the list. This is called a *free exchange*. Any other exchange of two consecutive items in the list incurs cost one and is called a *paid exchange*.

## 2.1.1 Full Cost Model

The full cost model developed by Sleator and Tarjan[2] is considered as the standard cost model for list accessing problem. According to this model, the cost for accessing a requested element is equal to the position of that element from the front of the list. For example, the cost of accessing an element 'x' at the $i^{th}$ position in the input list is equal to i.

## 2.1.2. Partial Cost Model

In partial cost model[3], the cost for accessing an element is the number of comparisons required for accessing the requested element in the input list. For example, the cost of accessing an element 'x' at the $i^{th}$ position in the input list is equal to i-1.

## 2.1.3 $P^d$ Cost Model

Manasse et. al[6] and Reingold et.al[7] introduced the $P^d$ cost model. In this model there are no free exchanges and each paid exchange costs d.

## 2.1.4. Centralized Cost Model

A cost model using doubly linked list, known as Centralised cost model, was developed by R. Mohanty et.al [6]. According to this cost model, access cost for a requested element is equal to its distance from the central element of the list. Free movement is moving the currently accessed element to any position forward or backward in the list towards the centre of the list with no cost. Paid movement is any exchanges other than the free movement. The cost incurred for paid movement is the distance between the elements to be exchanged.

## 2.2 List Accessing Algorithms

Many algorithms have been developed for the list accessing problem . The primitive algorithms are MTF, TRANSPOSE, and FC.

MTF: After accessing an element, the element is moved to the front of the list with no cost, without changing the relative order of the other elements in the list.

TRANSPOSE: After accessing an element of the list, it is exchanged with the immediately preceding element.

FREQUENCY COUNT: It maintains a frequency count for each element of the list, the count is initialised to zero. Then increase the count of an element by one whenever it is accessed and maintains the list so that the elements are in non-increasing order of their frequency count.

## 3. OUR PROPOSED COST MODEL

We have proposed a new cost model using the concept of buffering, look ahead and matching. In our proposed cost model, we have defined and used the following terminologies. A *List* is a sequence of unsorted distinct elements. *Request sequence* is a sequence of elements. *Visited list* is the portion of *the List* visited while searching for the requested element and it is marked by a pointer. We call this pointer *Visitor pointer*. *Matched elements* are the elements which results from parallel matching of Visited list and the next 'i' elements of the request sequence, where 'i' is the position of requested element in the input List. *Buffer* is the temporary memory which stores the matched elements. *Flag* is an extra bit given to the matched elements for identification purpose. *Flagged elements* are the elements, which are assigned a flag. The access of flagged item from Buffer costs 'i' if it is at the $i^{th}$ position in the buffer. The non flagged elements are accessed from the list by incurring access cost 'i' for an element in $i^{th}$ position of the List.

## 3.1. Assumptions

In our proposed cost model, we have assumed that the list is a singly linked list. For matching operation, we do the parallel matching. The matching cost is assigned as 'n' where 'n' is the number of matches. Maximum allowable Buffer size is given. If numbers of matched elements present are more than given buffer size, then the elements having higher 'i' values in the list ('i' is the position of element in the list) are placed in buffer. The list size is quite large. The visitor pointer always starts from starting of the list for each access in the list. In our proposed method, we use look ahead of 'i' from accessed element in request sequence. Here we know next 'i' elements from accessed elements in request sequence.

## 3.2. Proposed Cost Model

There are many existing cost models for list accessing problem. The previously existing cost models assume that the cost of rearrangement is zero. In our cost model, the list is not rearranged and we use buffer to store some elements for faster accessing. Our cost model assigns the processing cost of the request sequence as follows:

1. The cost of accessing an element x at the $i^{th}$ position in the input list and buffer is equal to i.

2. The matching cost is n, where n is the number of parallel matches .

3. As buffer space is limited, replacement occurs. The replacement cost is m where m is the number of elements replaced in buffer.

4. The processing cost of the request sequence is the sum of access cost, matching cost and replacement cost.

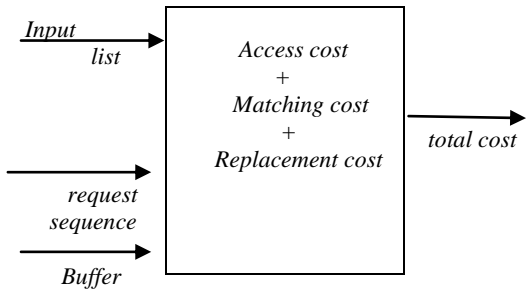

**Fig 1: Representation of proposed cost model**

## 3.3 Pseudo code For Our Proposed Algorithm

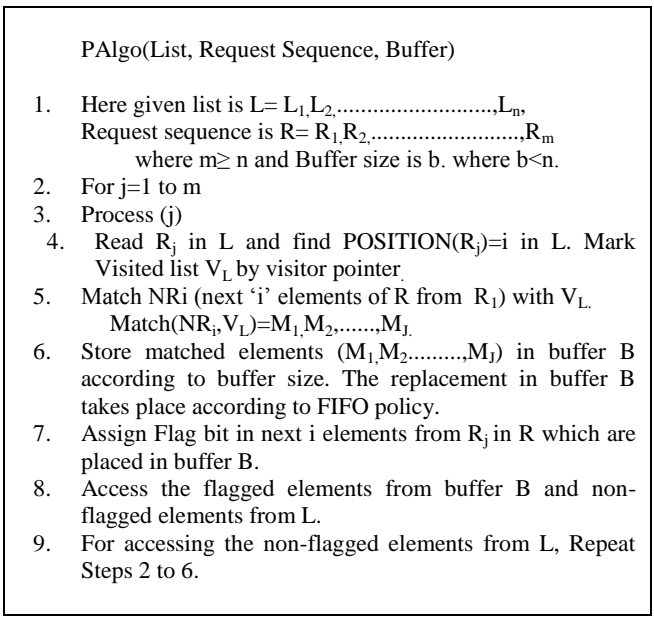

PAlgo(List, Request Sequence, Buffer)

1. Here given list is L= $L_1L_2$............................$L_n$,
   Request sequence is R= $R_1R_2$...........................$R_m$
   where m≥ n and Buffer size is b. where b<n.
2. For j=1 to m
3. Process (j)
4. Read $R_j$ in L and find POSITION($R_j$)=i in L. Mark Visited list $V_L$ by visitor pointer
5. Match NRi (next 'i' elements of R from $R_1$) with $V_L$
   Match($NR_i,V_L$)=$M_1,M_2,......,M_J$
6. Store matched elements ($M_1M_2$.........,$M_J$) in buffer B according to buffer size. The replacement in buffer B takes place according to FIFO policy.
7. Assign Flag bit in next i elements from $R_j$ in R which are placed in buffer B.
8. Access the flagged elements from buffer B and non-flagged elements from L.
9. For accessing the non-flagged elements from L, Repeat Steps 2 to 6.

*Illustration*

Given list L=A B C  D E F G H I  and Request sequence R= I E G D I E D  A B I and given buffer size is 3. We read I from L and access it with cost 9 as it is the ninth element in list. The visited list $V_L$ is marked by the visitor pointer and will be   A  B C D E F G H I. The elements of the look ahead 'i' are Next 9 elements of R from R I i.e, $NR_i$= E  G  I  E  D  A B  I. They are matched with  $V_L$ . Match($V_L,NR_i$)= E I . Matching cost is 2 as two matches occur in parallel matching . Then we store E I in Buffer B. Give flag to E I within look ahead 'i' i.e, within next 9 positions from I  in R. Next requested element is E, which is flagged. So access it from buffer with cost 1 as it is the first element in buffer. Then requested element is G, as it is non-flagged, it is accessed from L with cost 7. The visitor pointer mark the visited list $V_L$= A B C D E F G. Here $NR_i$=D I E D A B  I

Match($V_L$, $NR_i$) = D. Matching cost is 1. Already there are two elements in buffer and D is the third element. As buffer size is 3, no replacement needed.  Now buffer contains  E I D. Next element is D in R, as it is flagged, it is accessed from buffer with cost 3 as it is at the third position in buffer . The next request is

I, it is flagged so it is accessed from buffer with cost 2. Then E is requested. As it is flagged, the item is accessed from buffer with cost 1. Next request is for D, it is flagged. So it is accessed from buffer with cost 3. Next request is for A. It is non flagged, so it is accessed from list and $V_L$ marked by visitor pointer is $V_L$= A. $NR_i$ =B. No matches occur in parallel matching so buffer content remains the same. Next requested element is B. It is non-flagged. So it is accessed from  L with cost 2. The $V_L$ pointed by visitor pointer will be A B. $NR_i$= I. No matching occurs so buffer content remains same. Next request is I, it is flagged so it is accessed from buffer with cost 2. So the total cost for the above request sequence according to proposed cost model is 34.

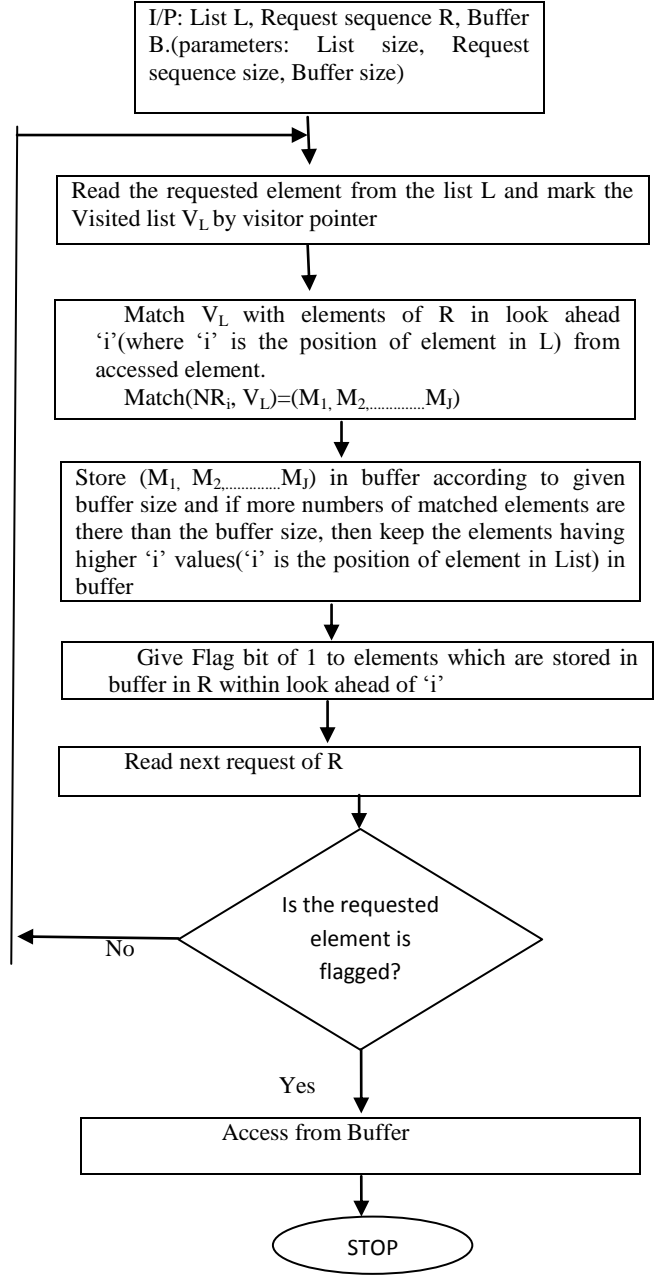

**Fig. 2  Flow Chart for Our Proposed Algorithm**

*Demonstration of Proposed Algorithm*

Given list is L= A  B  C  D  E  F  G  H I,  request sequence R= I E G  D  I  E  D  B  A  I. Buffer size is 3.

First request is for I, it is read from L with cost 9 and the visited list marked by visitor pointer is $V_L$= A  B  C  D  E  F  G  H I .POSITION(I) in L= 9. So next 9 elements from I in R is $NR_i$= E G  D  I  E  D    B  A  I. Match( $V_L$,$NR_i$) =E I with matching cost 2.

Given list is L= A  B  C  D  E  F  G  H I ,  request sequence R= I E  G  D  I  E  D  B  A  I    . Buffer size is 3.First request is for I, read it from L with cost 9 and the visited list marked by visitor pointer is $V_L$= A  B  C  D  E  F  G  H I .POSITION(I) in L= 9. So next 9 elements from I in R is $NR_i$= E  G  D  I  E  D    B  A  I    . Match( $V_L$,$NR_i$) =E I with matching cost 2.

Visitor pointer

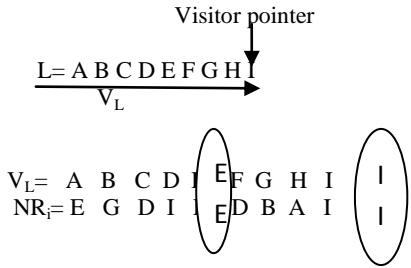

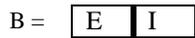

So store E I in buffer B. and give them flag in look ahead of 'i' i.e, in next 9 elements from I in R.

B =   [ E  |  I ]

The flagged elements in R are $NR_i$ = $E^I$ G D $I^I$ $E^I$ D  B A $I^I$

Next requested element in R is E, it is flagged so access it from buffer. The cost is 1 as it is at the first position in B. The next request is G. It is non-flagged.   POSITION(G)=7  in list.So access from L with cost 7. The $V_L$ marked by visitor pointer is A B C D E F G. The $NR_i$=D I E D  B  A I. Match($V_L$,$NR_i$)= D with matching cost 1.

Visitor pointer

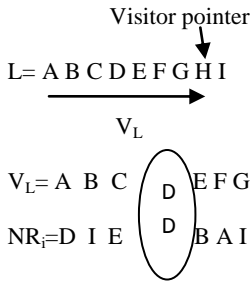

So store D in buffer B. now B=

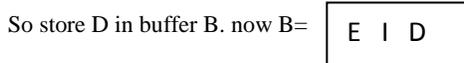

And give flag to E I D in R within look ahead of 'i' i.e, in next 7 elements from G in R.

$NR_i$=  $D^I$ $I^I$ $E^I$ $D^I$  B A $I^I$

Next request is D in R, it is flagged so access it from buffer with cost 3. Next request is for I. It is flagged so it is accessed from buffer with cost 2. Next element in R is E. It is flagged so it is accessed from buffer with cost 1. Next request is D. It is flagged and is accessed from buffer with cost 3. Next request is B, it is non-flagged. POSITION(B)=2 in list and it is accessed from L. The visited list marked by visitor pointer is  $V_L$=A B. $NR_i$= A I. Match($V_L$, $NR_i$)= A with matching cost 1.

Visitor pointer

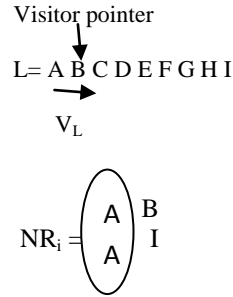

So, store A in buffer B. But buffer already contains        three elements so we have to replace one element by FIFO policy. So, E will be replaced by A with replacement cost 1

Now B=

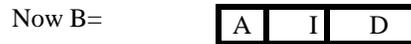

flag is given to A I D in $NR_L$
$NR_i$= $A^I$ $I^I$ .

Next requested element in R is A. It is flagged so access it from buffer B with cost 1. Next request is for I, as it is flagged ,access it from buffer with cost 2. The total cost for above request sequence according to proposed cost model is the sum of access cost, matching cost and replacement cost is 36

## 3.3  Comparison of proposed cost  model with standard cost model

We have performed experiment by implementing our proposed algorithm using AMR cost model and MTF algorithm using full cost model. We have calculated total cost of each method for different list configuration and request sequence. We have compared the total cost of our proposed algorithm using AMR cost model with the MTF algorithm using full cost model. Here we have observed  that our algorithm  using AMR model performs better than the MTF algorithm as shown in table-1. Figure-3 shows the comparative cost of MTF algorithm using Full cost model and the new proposed algorithm using AMR cost model.

From the analysis, we have observed that when the elements having higher 'i' values (where 'i' is the position of element in the input list) are the buffered elements, then the gain is more. When the buffered elements are frequently present in the request sequence, then this cost model gives significant gain. When no matching occurs and the elements having higher 'i' values are repeatedly present in the request sequence, then it performs worst.

## 4.  CONCLUSION AND FUTURE WORK

In this paper, we have presented new cost model using singly linked list data structure, which considers access cost, matching cost, and replacement cost . Here we have proposed one method which involves the matching, buffering, look ahead and flag bit. This method calculates the processing cost of request sequence using our proposed cost model. We have compared our work with MTF which calculates the cost using standard cost model.

| Sl. No | Request sequence size | Request sequence | Access Cost of MTF using Standard cost model | Cost Using Proposed Cost Model |
|---|---|---|---|---|
| 1 | 11 | K J I H G F E D C B A | 121 | 66 |
| 2 | 19 | H B B A A E E G H J J A J C E H B H H | 80 | 58 |
| 3 | 20 | G B A D D G F D B A A H G C B A H G F H | 87 | 48 |
| 4 | 21 | K B B A I I J K I I J K H I B B A A K J I | 92 | 54 |
| 5 | 22 | J B J J J E J G G E J I J I J I J I J I I I | 66 | 64 |
| 6 | 23 | K J A B A H K B H H J K I I J D C J K B K I | 125 | 63 |

We have analysed our work and given the cases when it performs best and worst.

For matching operation, here we have used the parallel matching technique. By using improved matching techniques the proposed cost model can be improved. In our proposed cost model, we have assumed that the buffer size is given and it is static. In future work, it can be made dynamic for improve the efficiency. Different list update algorithms can be developed by using this cost model. For replacement in buffer, we have used FIFO policy; other advanced paging policy can be included for extension.

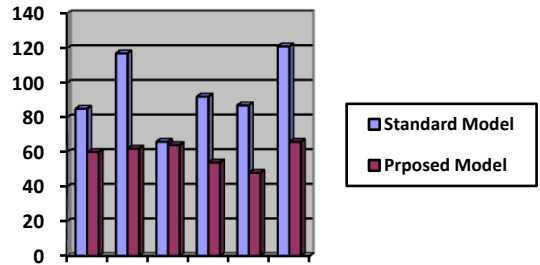

**Fig. 3 Comparison of AMR Model with Standard Cost Model**